# Generative AI in Science: Applications, Challenges, and Emerging Questions

Ryan Harries
*School of Social Sciences*
*University of Manchester*
Manchester, UK
ORCID: 0000-0002-1866-0985

Cornelia Lawson
*Manchester Institute of Innovation Research*
*University of Manchester*
Manchester, UK
ORCID: 0000-0002-1262-5142

Philip Shapira +
*Manchester Institute of Innovation Research*
*University of Manchester*
Manchester, UK
*and*
*The Jimmy and Rosalynn Carter School of Public Policy*
*Georgia Institute of Technology*
Atlanta, USA
ORCID: 0000-0003-2488-5985

*Abstract*— This paper examines the impact of Generative Artificial Intelligence (GenAI) on scientific practices, conducting a qualitative review of selected literature to explore its applications, benefits, and challenges. The review draws on the OpenAlex publication database, using a Boolean search approach to identify scientific literature related to GenAI (including large language models and ChatGPT). Thirty-nine highly cited papers and commentaries are reviewed and qualitatively coded. Results are categorized by GenAI applications in science, scientific writing, medical practice, and education and training. The analysis finds that while there is a rapid adoption of GenAI in science and science practice, its long-term implications remain unclear, with ongoing uncertainties about its use and governance. The study provides early insights into GenAI's growing role in science and identifies questions for future research in this evolving field.

## I. Introduction

Generative Artificial Intelligence (GenAI) marks a significant advancement in AI technology, characterized by its encompassing capabilities and widespread accessibility. Across all fields of science, researchers are increasingly leveraging GenAI to enhance research methodologies, streamline the writing of scientific communications, and innovate in testing, diagnostics, and training. However, the rapid adoption of GenAI in science has also triggered debate regarding its trustworthiness, reproducibility, ethical implications, and challenges for governance and equity. While examples demonstrate the promise of these tools, the full scope of GenAI's current and future roles in science remains uncertain, reflecting both the speed of its development and the complexity of its integration into established and new scientific practices.

This paper offers a qualitative review of literature and commentary on GenAI's early impacts on science, analyzing perspectives from high-impact publications through thematic analysis. By categorizing findings into key themes, the study identifies critical areas of transformation and tension surrounding GenAI's deployment. Applications of GenAI are explored across research processes, scientific writing, medical practice, and educational contexts.

After discussing GenAI and its emergence, the methodology used in the review is presented. This is followed by a thematic analysis and discussion of findings and then by conclusions. Questions for future research are also considered.

## II. GenAI Emergence

GenAI is a development from earlier Artificial Intelligence (AI) models that use both supervised and unsupervised learning and whose algorithms are constrained by the limited size of the data they can process. However, newer GenAI models use much larger data sets and, with the development of 'convoluted neural networks' and 'recurrent neural networks', can analyze images, texts and video (Dwivedi et al., 2023). These GenAI models are trained on extensive data sets that can create new data and/or context (such as text, images or music) based on patterns learned from preexisting information (Pavlik, 2023; Ray, 2023). For example, Pavlik states that "a generative model might be trained on a dataset of images of faces and then be able to generate new, previously unseen faces that look realistic" (2023, p. 87). Other examples include generating music, painting, poetry and deepfakes (Floridi & Chiriatti, 2020). GenAI can generate outputs and/or content that closely resemble human-created content in a wide array of contexts (Ray, 2023).

A key feature regarding the applicability of GenAI over a variety of contexts is its integration with large language models (LLMs). LLMs refer "to systems which are trained on string prediction tasks: that is, predicting the likelihood of a token (character, word or string) given either its preceding context or (in bidirectional and masked LMs) its surrounding context" (Bender et al., 2021, p. 611). Newer versions or two staged/bidirectional LLMs have two major advantages, namely 1) they are able to process much larger data sets and 2) these larger amounts of data are able to be processed without a great deal of human manual input – hence, newer LLMs can be initially trained on quite a small data set with human input and then have a much greater scalability and accuracy without the need for a great deal of human input in these later stages (Shen et al., 2023). This allows for conversational-like or Generative Pre-Trained Transformer (GPT) models that have emerged out of the field of neurolinguistic programming, facilitation the development of tools that use machine learning to better comprehend human language in order to produce and reply to text (Ray, 2023; Salvagno et al., 2023) of which ChatGPT is the most notable.

While the first iteration of ChatGPT (software intending to mimic human-like discussions with users) emerged in 2018, it was after its public release in November 2022 that it quickly experienced massive growth in its user base and applications (Floridi & Chiriatti, 2020; Lo, 2023). ChatGPT can perform at least as well as humans on a number of both professional and academic tests (Ray, 2023). For example, ChatGPT articulates its responses to such a level that it has been able to pass medical exams (Nature, 2023). This is largely a result of ChatGPT being able to use advanced language processing with a wide range of language applications such as translation, text summarization, being able to answer questions on a wide range of topics, editing the style and grammar of text, as well generating text for users, for example composing emails (Cotton et al., 2023; Salvagno et al., 2023). ChatGPT has been deployed in the business sector for multiple applications, including customer service, and used to aid various forms of content creation (Cotton et al., 2023; Floridi & Chiriatti, 2020; Ray, 2023).

From the view of science, while precursors to – and prototypes of – GenAI models were available in the 2010s and the early years of the 2020s, it was the launch of OpenAI's ChatGPT in November 2022 that greatly boosted the capabilities of and accessibility to GenAI (Bengesi et al., 2024; Lawlor & Change, 2024). In was in this year that scientific interest in GenAI began to take off – an interest that grew exponentially from 2023 as multiple other GenAI models plus enhanced versions of ChatGPT were launched. GenAI and GPTs have multiple uses and implications in scientific fields. These technologies have been used within biological research, for example in de novo drug design (Gupta et al., 2017; Merk et al., 2018; Rives et al., 2021) and the development of new protein structures (Van Noorden & Perkel, 2023). Within the medical field it has also been used as a tool for generating clinical notes and analyzing medical images (Lee et al., 2023) and suggesting medical diagnoses (Van Noorden & Perkel, 2023). Among research practices, ChatGPT has been used to aid with statistical analysis (Cascella et al., 2023) and also has the potential to aid in several research design practices such as identifying potential research questions, selecting suitable methodologies and aid in scoping the existing state of the field and fabrication of a literature review (Dave et al., 2023; Salvagno et al., 2023). Given that GPTs can also aid with the writing and publication process, this poses questions for scientific integrity and authorship (Stokel-Walker, 2023).

While we recognize that GenAI is still at early stages of development and application, it is evident that there are significant implications for science with the emergence of new GenAI technologies. The aim of this paper is to probe these developments by unpacking the plethora of applications affecting science and science practices and by discussing concerns raised in the literature and imagined possibilities of the use of such technologies within scientific fields.

## III. Methodology

The study was guided by an overarching research question which asked, "What are the applications and implications of the use of GenAI in science?" In the paper, we address this question by conducting a qualitative in-depth review of selected papers and communications across scientific fields to explore examples and viewpoints on how GenAI is currently influencing science, science practice, and science education and what might be its potential impacts going forward.

The methodological approach pursued in the study was, first, to identify the population of relevant publications, and select those GenAI studies to be included for review; second, to undertake qualitative thematic review and coding. Each of these procedures involved several methodological steps, as explained in the balance of this section.

The database used to identify the population of relevant publications is OpenAlex. This open source data set provides comprehensive scientific publication metadata (Priem et al., 2022). OpenAlex collates many types of scientific works including journal articles, book chapters, preprints (important in computer science fields), reviews, editorials and commentaries, and books. Our dataset is sourced from the Open Alex January 2024 snapshot, which encompasses over 250 million works.

To identify GenAI works in Open Alex, we used a Boolean search approach that *included* terms applicable for Generative AI but also *excluded* certain terms that would generate false positives. This search approach was developed and validated as part of the Project on Innovations in the Lab: Leveraging Transformations in Science at the Manchester

Institute of Innovation Research (Ding et al., 2024). The search approach was as follows:

("generative pre-trained transformer" or "Generative language model" or "Generative large language model" or "generative artificial intelligence" or "Generative pre-trained language model" or "chatgpt" or "openai large language model" or "openai gpt" or "GenAI" or "Generative AI" or "generative AI" or "GPT-1" or "GPT-2" or "GPT-3" or "GPT-4") NOT ("Glutamate Pyruvate Transaminase" or "Glutamate-Pyruvate Transaminase" or "glutamic-pyruvic transaminase" or "N-p-tolyl-D-glucosylamine" or "UDP-GlcNAc:dolichol-P GlcNAc-1-P transferase" or "Goniopora toxin" or "General Professional Training" or "guinea-pig isolated trachea" or "gradational psychosomatic treatment" or "General Purpose Technology" or "Glc6P/phosphate translocator" or "gemcitabine chemotherapy" or "gemcitabine-concurrent proton radiotherapy" or "Gas production technique" or "UDP-GlcNAc:dolichyl-phosphateN-acetylglucosamine 1-phosphate transferase")

The search approach was applied across titles, abstracts, and full text and encompassed all 19 of Open Alex's root concepts (or research fields) comprising (in alphabetical order): art, biology, business, chemistry, computer science, economics, engineering, environmental science, geography, geology, history, materials science, mathematics, medicine, philosophy, physics, political science, psychology, and sociology. After de-duplication and other preprocessing, 13,661 GenAI publication records were identified, covering the period 2017 to 2023. Of these, 91.2% were published in 2023, with 2.2% and 5.0% in 2021 and 2022 respectively.

In contrast to quantitative bibliometric studies which analyze large numbers of publications but typically focus on machine-readable elements (e.g., titles, abstracts, key words, affiliations, and citations), a key aim in the present study was to read and qualitatively code the full text of contributions across the range of scientific fields. We sought to identify nuances in how scientists were using GenAI in science and how science itself was or might be affected by the diffusion of GenAI. To make human reading feasible, we needed to select a manageable number of publications. Our approach focused on identifying publications and commentaries with 100 or more citations as indicated in the Open Alex January 2024 snapshot. Citations represent a quantifiable way to identify papers that have been found by peers to be useful and relevant, although we note that field size affects citation accrual and that there are other issues with citations including time lags. The citation selection method of a fast emerging topic privileges disciplines where papers can be rapidly published and able to garner relatively large number of citations. In sum, while readily implemented, the limitations of our approach to sample selection should be kept in mind.

After removing one non-relevant paper, thirty-nine papers were identified for qualitative review (see Appendix for full list). Citations to these papers ranged from 100 to 1044 (median 186). Of these papers, 15 were published in journals in medical and health sciences, 9 in natural sciences (including computer sciences), 8 in social sciences (including education), and 7 in multidisciplinary outlets (such as *Nature*, *Science*, and *PNAS*). The sample is primarily comprised of journal papers but also includes two proceedings papers and one repository paper.

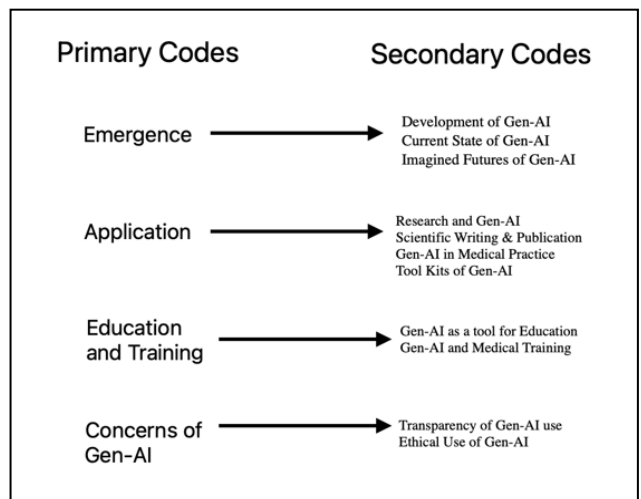

*Figure 1. Outline of Thematic Coding Scheme*

The selected publications were downloaded, thematically analyzed and qualitatively coded using NVivo 12 software. From this, four higher-level primary codes were identified which consisted of 1) Emergence 2) Application 3) Education and Training and 4) Concerns of GenAI. Under these codes, eleven secondary codes were identified (Figure 1). Emergence secondary codes consisted of the Development of GenAI, Current State of GenAI and Imagined Futures of GenAI. Applications subcodes were as follows: Research and GenAI, Scientific Writing & Publication, GenAI in Medical Practice, and Tool Kits of GenAI. Education and Training had two subcodes, namely, GenAI as a tool for Education and GenAI for Medical Training. The final set of subcodes under Concerns of GenAI consisted of Transparency of GenAI use and Ethical Use of GenAI.

IV. ANALYSIS

The qualitative analysis of the reviewed publications found that GenAI, LLM's and related models such as ChatGPT are impacting – or are anticipated to impact – scientific fields in varied ways. Through applying thematic analysis to our selected literature set, using the codes noted in the previous section, we were able to discern specific insights about where and how scientists were using these new technologies. While the literature indicated multiple applications, it also identified many observations speaking to concerns as well as the potential for GenAI use in science and science practice. The following sections present our results

*A. Applications of GenAI in Research*

In the papers we reviewed, examples and suggestions were presented that GenAI is applicable in multiple ways to various aspects of scientific research. For example, Dave et al (2023) illustrate that ChatGPT can be quite useful in the initial stages of research as it can be used as an advanced 'search engine' that could aid in the processes of conducting a literature review, where the tool could aid in the appropriate selection of articles and write up of the review which would allow the researcher more time to focus on the selection of applicable methodologies and doing the actual research.

Other authors say that ChatGPT can assist with research summarization as well as analysis of the work providing insight into possible research questions (Lee et al., 2023). Additionally, Casecella et al (2023) suggest the use of this tool could in translating statistical analysis and aiding the development of appropriate research models. This alludes to the use of GenAI tools as a means of translating science among researchers as well as to the broader public.

In addition to its uses in science practice and communication, examples were presented of the use of GenAI technology deep into scientific research processes. Examples are found in de-novo drug design (Merk et al., 2018; Rives et al., 2021), where there is discussion of the benefits of the technology in testing biological compounds, lessening (if not replacing) the need for physical research, and in facilitating unsupervised machine learning leading to faster lead times for research (Gupta et al., 2017).

Also discussed in the literature is the integration of GenAI across the stages of the research process through to the publication of results. As Dowling & Lucy state: "What we have shown in this study is important. ChatGPT can generate, even in its basic state, plausible-seeming research studies for well-ranked journals" (2023, p. 5).

However, the analysis of the literature set finds that the use of GenAI technology is also marked by a series of concerns. Major issues reported are the sustainable, equitable and transparent use of the technologies within research (Bender et al., 2021; Dave et al., 2023; Kasneci et al., 2023). The sustainable use of GenAI is problematic as the energy requirements to train and run models are high. For example training a single BERT model may use as much as a trans-American flight (Bender et al., 2021). There also remains the issue of scalability and upgrading of energy-efficient hardware and data centers to deal with the high computational load of cloud networks (Cooper, 2023). This would indicate a relative concern for the inequalities that may develop between high-income countries and low-income countries as illustrated by Dave et al "as with other technologies, high-income nations and privileged academics are likely to eventually discover methods to leverage LLMs in ways that advance their research while exacerbating disparities" (2023, p. 3).

Beyond the disparities between low-income and high-income countries, there are further concerns that increased use of these technologies will lead to more sophisticated algorithms and modules that will be trained on larger data sets creating an increased likelihood for misuse and manipulation (Dave et al., 2023). These concerns are compounded by ChatGPT being referred to as a Blackbox technology (Sallam, 2023) where users and not entirely sure how and why they receive the outputs as well as how private companies are using individuals' data to train models. There are also concerns over the accuracy of outputs from these models, with hallucination effects cited (Sallam, 2023) where outputs from models are made up, inaccurate or without coherent referencing.

GenAI models raise other concerns related to the epistemological foundations of science and to power and equity. For example Cooper (2023) worries that ChatGPT runs the risk of being the 'ultimate epistemological authority' which begs the questions about whose voices are ignored in the process and what biases are assumed. There are further concerns over GenAI technology following established patterns in modernity where newer technologies tend to favor the most privileged within society (Bender et al., 2021). These observations underscore worries about the extent to which GenAI poses risks to the integrity and openness of scientific research as well as ethics surrounding research particularly relating to sustainability, integrity and equality.

Overall, and at the current points of development and diffusion of GenAI, the literature is disjointed with regard to the role that these technologies will play within the research process. On the one hand, there are promises with some examples of how GenAI will boost research performance and scientific discovery. On the other hand, there are concerns about a series of negative implications, also with some examples. It is not apparent (at least at present) that GenAI technologies will reach the stage where they no longer need human intervention to carry out research projects, but it does seem evident that with the new GenAI applications that are now available, human roles in science and the processes of science are changing. The balance between incremental and fundamental change is unclear. Moreover, while many adverse aspects of GenAI applications are raised, as yet there is no consensus on how these can be addressed and by whom.

*B. GenAI in Scientific Writing and Publication*

Scientific publication is pivotal in science and scientific advancement, being crucial to the validation and communication of scientific discoveries and results. Publication is also vital in scientists' career development. Additionally, publications are used in science metrics and organizational science performance assessments. The uptake of GenAI and related models in science, particularly its advanced language processing capabilities, has sparked much debate surrounding the use of these technologies in the writing and publication process.

ChatGPT has been explicitly used in the scientific publication process. It is listed as an author on at least 12 pre-prints (Stokel-Walker, 2023), as a co-author on a paper on nursing education (O'Connor, 2023), has written a full physics paper and used to outline a paper (Dowling & Lucey, 2023) and has been listed as a tool in a number of the papers found within our literature review (Biswas, 2023; Ray, 2023). In short, there has been an uptick in use of ChatGPT in the writing process. However, these documented instances are just a fraction of the more frequent but less visible uses of ChatGPT to assist scientific writing but without necessarily acknowledging this.

At the most basic level, the technology is able to assist in the editorial process, identifying basic grammatical errors, aid with overcoming writers block, help the translation and writing of papers that are not written in English, and even help with image and figure manipulation (Cotton et al., 2023; Kasneci et al., 2023; Shen et al., 2023). However, there are many further applications of these technologies for scientific writing as well as the publications process. It is expected that "this technology will evolve to the point that it can design experiments, write and complete manuscripts, conduct peer review and support editorial decisions to accept or reject manuscripts" (van Dis et al., 2023, p. 224). This has generated debates over the use of the technology within the scientific writing and publication process, with concerns about plagiarism, authorship rights, scientific integrity and the detection of these technologies within written pieces. Thorp offers good insight into these questions:

> "The scientific record is ultimately one of the human endeavour of struggling with important questions. Machines play an important role, but as tools for the people posing the hypotheses, designing the experiments, and making sense of the results. Ultimately the product must come from—and be expressed by—the wonderful computer in our heads" (2023, p. 313)

In this debate over publication, we are reminded that authors are responsible and accountable for the legitimacy and truthfulness of their work in publication. This is an aspect that ChatGPT cannot do, reiterating the point that it cannot be an author (Stokel-Walker, 2023). Others argue that GPTs are more akin to other research software such as 'Microsoft Word' or 'Python' and hence its use needs to be listed in the methodology or acknowledgement sections (Cooper, 2023). For some, this poses issues of scientific integrity particularly concerning the use of such tools for image manipulation and figure generation, and it recommended that publication guidelines be modified to state that the use of GenAI in these areas would constitute scientific misconduct (Thorp, 2023).

It appears that the use of GenAI tools is still a grey area and there are major concerns over plagiarism and the ability of publishers, editors, and reviewers to detect work that has been submitted using GenAI. The literature notes the concerning ability of these tools to bypass current plagiarism detectors (Dave et al., 2023; Lo, 2023) and to pass through peer review processes. As Shen et al. (2023, p.3) observe, "early data suggests that while peer reviewers will likely be able to identify the most egregious offenders and have a high suspicion for others, better written or fabricated articles that undergo subsequent polishing may fool initial screening and/or inexperienced peer reviewers."

A particular issue in using ChatGPT in scientific writing is the ability to check the accuracy of the text and references it generates. Kitamura (2023) illustrates that ChatGPT often generates inaccuracies and the use of the technology is only helpful if researchers can ensure accuracy of the work it generates. The author also points out that software such as ChatGPT tends to generate text in a repetitive fashion which may limit authors' creativity. In other words, an overreliance on the GPTs could lead to a lack of original human thought when writing.

The literature points to a paradox in which these technologies could aid with writing and thought as well as hinder it. Scientific writing could be transformed through GPT idea generation as well as editing and translation assistance. On the other end of the spectrum are concerns that the software poses a serious risk to academic integrity, responsible authorship and detection of misuse. Again, understanding of the full range of impacts generated by the use of GenAI technologies in the process of scientific writing and publication is still somewhat speculative. This remains a topic for future research.

*C. GenAI in Medical Practice*

A major area of focus within the selected scientific literature was on the current and potential applications of GenAI within the field of medicine. We those devote additional attention to applications of GenAI within and around the clinical setting and its potential to change medical practices.

One suggested application of GenAI in the clinical setting is pre-authorization (Shen et al., 2023). Here, a possible strength of ChatGPT within the clinical context is that it could speed up the process of gaining authorization from insurance companies and thus save doctors time. This is due to the potential of ChatGPT to scan patients' history and relate it to the current ICD-10 (International Statistical Classification of Diseases and Related Health Problems) and thus, could draft letters explaining why patients would need certain procedures.

The use of GPTs within the clinical context is not limited to assurance approval, with other authors suggesting several other methods. Lee et al are confident that even in ChatGPT's current state it could accurately "be used for generating clinical notes by just providing raw information, such as daily summaries and discharge summaries, which could save time and increase accuracy" (2023, p. 1235). They highlight the potential of ChatGPT to better communicate with patients when they leave the clinical setting, with tools such as Medical Chatbots providing personalized or tailored responses to patients over the course of their treatments.

Cascella et al suggest that one of ChatGPT's strongest attributes is its "ability to summarize information, although sometimes imprecise, using technical language for communication among clinics as well as plain language for communication with patients and their families" (2023, p. 32). It appears that ChatGPT could be a useful tool for summarization and communication within the clinical setting and could enhance the doctor-patient relationship with faster and clearer communication.

GPTs can also be used to aid doctors when assessing patients in several suggested ways. Shen et al (2023) suggest that ChatGPT can assist in reviewing and summarizing the patients' health records which could potentially aid in the diagnostic process. Furthermore, GPTs have been shown to help radiologists with image diagnostics. Chatbots within the healthcare system have been suggested to potentially aid in faster triage within the clinical space (Patel & Lam, 2023).

While there are promising benefits to adopting GenAI technology in medical practice, these do not come without concerns. A major issue is that there are currently no laws that govern the use of the technology within the field. This poses issues such as concerns over patients' data as well as accountability for misdiagnosis due to using the tools (Biswas, 2023). While there are many opportunities for automation with these technologies (Patel & Lam, 2023) given that the medial field has a high stake in ensuring accuracy and reducing risk to patients, this opens the field up to a wide array of legal ramifications if the technology is not governed properly. Moreover, medical practice is not free from multiple concerns about the use of GenAI technology such as hallucination effects, bias, transparency, and accuracy. The literature however fails to largely account for who is responsible for the variety of concerns facing the uptake of GenAI technology outside of legal frameworks. A possible emerging research question then could be: how would the literature suggest 'responsibilities' for a field such as medicine, outside of just a legal framework?

*D. GenAI in Education and Training*

A further topic of emphasis within our selected literature set was on GenAI's implications for education and training in science, particularly in medicine. GenAI promises to create a more personalized learning experience for trainees, reduce administrative tasks and potentially provide the structure for faster turnaround times during assessment periods.

The use of various 'intelligent' systems for the purposes of teaching is not a new phenomenon, for example, systems to provide one-to-one tutoring such as intelligent tutoring systems (ITS) have been around for some time now (Zhai, 2022). However, GenAI is positioned by various authors as having a great capacity for more personalized learning experiences through the use of ChatBots and adaptive and personalized learning from GenAI systems (Cotton et al., 2023; Kasneci et al., 2023; O'Connor, 2023; Zhai, 2022). It could provide various benefits such as adaptive lessons to individuals skill levels that are progressively challenging while using various resources which may better suit student needs and preferences (Zhai, 2022). Beyond teaching, GenAI technology could be used to aid with academic timetabling and the creation of assessments (Zhai, 2022) and even potentially help with the marking of assessments including in medical fields (Dave et al., 2023; O'Connor, 2023).

GenAI technology raises some concerns, including the quality of teaching as well as the quality of assessment administered. O'Connor offers good insight into this when speaking about GenAI use in nursing "AI software cannot substitute for the knowledge, skills, and critical thinking that nursing students must develop and demonstrate in their university assessments. The use of AI software could lead to an erosion of the quality of the assessment, with implications for the development of the nursing profession" (2023, p. 1). This area is also not free from the other concerns posed in earlier sections, such as the ethical use of data (of both students and teachers), copywriting issues, costs of training and setting up these models for effective use in education as well as the sustainable and equitable use of these in the education field (Kasneci et al., 2023). This raises questions for further research, including who is responsible for ensuring the quality of assessments within scientific education and how are these technologies going to be rolled out equitably and sustainably in the future?

## V. Conclusions

Drawing on a qualitative literature review, the paper has described various applications of GenAI in science and science practice and discussed issues raised by these applications. This paper is an outcome of work in progress. Many questions for further research lines are raised, of which key emerging research questions are listed in Table 1.

What we can say at this point is that the emergence of new GenAI tools has generated much debate over its potential in science and its implications. There has been an exponentially increasing use of various GenAI tools in science. Yet, notwithstanding examples and a growing number of studies, the full extent of how GenAI is being used in science and science practice, and how it will evolve and be used in the future, remains somewhat unclear. The emergence and diffusion of GenAI in science is occurring so rapidly that science is itself not certain about the technology's use, boundaries, and implications. Issues of ethics, trust, sustainability, responsibility, equity, and governance are raised, without as yet clarity or consensus as to how these might be resolved. Going forward, the review observes expectations of continued growth of applications of GenAI in science and science practice, regardless of the many uncertainties that surround it.

*Table 1. Generative AI in Science: Emerging Research Questions*

1) How do scientific practices involving the use of GenAI change the production of knowledge, particularly relating to theory?
2) What human tasks in to science will GenAI reduce? Augment? Add?
3) Will GenAI reach the point where it can undertake research without the need for human input? Or, will GenAI remain a tool that inevitably requires human input?
4) What cybersecurity issues are raised with the increasing use of GenAI in scientific fields?
5) If GenAI can be used for nefarious scientific purposes, who can this be tracked and addressed?
6) If legal concerns are raised about the use of GenAI in science, i.e. in medicine, how should these be dealt with?
7) Many concerns are raised about the ethical and responsible use of GenAI in science. How should these be addressed?
8) What are the differences in the uptake of GenAI in science between early career researchers and senior academics? And by other characteristics of the scientific labor force (e.g., gender, ethnicity, discipline, qualifications, institutional type, country)?
9) How can issues raised by the use of GenAI in scientific publishing be addressed?
10) How is GenAI changing practices in scientific education and training, and with what implications?

Acknowledgments

This work was supported in part by the Project on Innovations in the Lab: Leveraging Transformations in Science, FHUMS Large Collaborative Grant, University of Manchester, UK (CL, PS). RH was supported through the Small Grant Scheme of the Manchester Institute of Innovation Research (MIOIR), Alliance Manchester Business School, The University of Manchester. The authors acknowledge assistance in the bibliometric search from Liangping Ding. Additional feedback was provided by seminar participants of the AI & Policy Discussion Group (MIOIR and the School of Public Policy, Georgia Institute of Technology) and by Julie Jebsen.

# REFERENCES

Bender, E. M., Gebru, T., McMillan-Major, A., & Shmitchell, S. (2021). On the Dangers of Stochastic Parrots, Proceedings of the 2021 ACM Conference on Fairness, Accountability, and Transparency, 610-623. https://doi.org/10.1145/3442188.3445922

Biswas, S. (2023). ChatGPT and the Future of Medical Writing. Radiology, 307(2). https://doi.org/10.1148/radiol.223312

Bengesi, S., El-Sayed, H., Sarker, M.K., Houkpati, Y., Irungu J., & Oladunni, T. (2024). Advancements in Generative AI: A Comprehensive Review of GANs, GPT, Autoencoders, Diffusion Model, and Transformers. IEEE Access, 12, 69812-69837. doi:10.1109/ACCESS.2024.3397775

Cascella, M., Montomoli, J., Bellini, V., & Bignami, E. (2023). Evaluating the Feasibility of ChatGPT in Healthcare: An Analysis of Multiple Clinical and Research Scenarios. Journal of Medical Systems, 47(1). https://doi.org/10.1007/s10916-023-01925-4

Cooper, G. (2023). Examining Science Education in ChatGPT: An Exploratory Study of Generative Artificial Intelligence. Journal of Science Education and Technology, 32(3), 444-452. https://doi.org/10.1007/s10956-023-10039-y

Cotton, D. R. E., Cotton, P. A., & Shipway, J. R. (2023). Chatting and cheating: Ensuring academic integrity in the era of ChatGPT. Innovations in Education and Teaching International, 61(2), 228-239. https://doi.org/10.1080/14703297.2023.2190148

Dave, T., Athaluri, S. A., & Singh, S. (2023). ChatGPT in medicine: an overview of its applications, advantages, limitations, future prospects, and ethical considerations. Frontiers in Artificial Intelligence, 6. https://doi.org/10.3389/frai.2023.1169595

Ding, L., Lawson, C., & Shapira, P. (2024). Rise of Generative Artificial Intelligence in Science. arXiv. https://arxiv.org/abs/2412.20960.

Dowling, M., & Lucey, B. (2023). ChatGPT for (Finance) research: The Bananarama Conjecture. Finance Research Letters, 53. https://doi.org/10.1016/j.frl.2023.103662

Dwivedi, Y. K., Kshetri, N., Hughes, L., Slade, E. L., Jeyaraj, A., Kar, A. K., Baabdullah, A. M., Koohang, A., Raghavan, V., Ahuja, M., Albanna, H., Albashrawi, M. A., Al-Busaidi, A. S., Balakrishnan, J., Barlette, Y., Basu, S., Bose, I., Brooks, L., Buhalis, D., Carter, L., Chowdhury, S., Crick, T., Cunningham, S. W., Davies, G. H., Davison, R. M., Dé, R., Dennehy, D., Duan, Y., Dubey, R., Dwivedi, R., Edwards, J. S., Flavián, C., Gauld, R., Grover, V., Hu, M.-C., Janssen, M., Jones, P., Junglas, I., Khorana, S., Kraus, S., Larsen, K. R., Latreille, P., Laumer, S., Malik, F. T., Mardani, A., Mariani, M., Mithas, S., Mogaji, E., Nord, J. H., O'Connor, S., Okumus, F., Pagani, M., Pandey, N., Papagiannidis, S., Pappas, I. O., Pathak, N., Pries-Heje, J., Raman, R., Rana, N. P., Rehm, S.-V., Ribeiro-Navarrete, S., Richter, A., Rowe, F., Sarker, S., Stahl, B. C., Tiwari, M. K., van der Aalst, W., Venkatesh, V., Viglia, G., Wade, M., Walton, P., Wirtz, J., & Wright, R. (2023). "So what if ChatGPT wrote it?" Multidisciplinary perspectives on opportunities, challenges and implications of generative conversational AI for research, practice and policy. *International Journal of Information Management*, 71. https://doi.org/10.1016/j.ijinfomgt.2023.102642

Floridi, L., & Chiriatti, M. (2020). GPT-3: Its Nature, Scope, Limits, and Consequences. Minds and Machines, 30(4), 681-694. https://doi.org/10.1007/s11023-020-09548-1

Gupta, A., Müller, A. T., Huisman, B. J. H., Fuchs, J. A., Schneider, P., & Schneider, G. (2017). Generative Recurrent Networks for De Novo Drug Design. Molecular Informatics, 37(1-2). https://doi.org/10.1002/minf.201700111

Kasneci, E., Sessler, K., Küchemann, S., Bannert, M., Dementieva, D., Fischer, F., Gasser, U., Groh, G., Günnemann, S., Hüllermeier, E., Krusche, S., Kutyniok, G., Michaeli, T., Nerdel, C., Pfeffer, J., Poquet, O., Sailer, M., Schmidt, A., Seidel, T., . . . Kasneci, G. (2023). ChatGPT for good? On opportunities and challenges of large language models for education. Learning and Individual Differences, 103. https://doi.org/10.1016/j.lindif.2023.102274

Kitamura, F. C. (2023). ChatGPT Is Shaping the Future of Medical Writing But Still Requires Human Judgment. Radiology, 307(2), e230171. https://doi.org/10.1148/radiol.230171

Lawlor, P., & Change, J. (2024). The rise of generative AI: A timeline of breakthrough innovations. Qualcomm, onQblog. https://www.qualcomm.com/news/onq/2024/02/the-rise-of-generative-ai-timeline-of-breakthrough-innovations

Lee, P., Bubeck, S., & Petro, J. (2023). Benefits, Limits, and Risks of GPT-4 as an AI Chatbot for Medicine. New England Journal of Medicine, 388(13), 1233-1239. https://doi.org/doi:10.1056/NEJMsr2214184

Lo, C. K. (2023). What Is the Impact of ChatGPT on Education? A Rapid Review of the Literature. Education Sciences, 13(4). https://doi.org/10.3390/educsci13040410

Merk, D., Friedrich, L., Grisoni, F., & Schneider, G. (2018). De Novo Design of Bioactive Small Molecules by Artificial Intelligence. Molecular Informatics, 37(1-2). https://doi.org/10.1002/minf.201700153

Nature. (2023). Tools such as ChatGPT threaten transparent science; here are our ground rules for their use. Nature, 613(7945), 612. https://doi.org/10.1038/d41586-023-00191-1

O'Connor, S. (2023). Open artificial intelligence platforms in nursing education: Tools for academic progress or abuse? Nurse Educ Pract, 66, 103537. https://doi.org/10.1016/j.nepr.2022.103537

Patel, S. B., & Lam, K. (2023). ChatGPT: the future of discharge summaries? The Lancet Digital Health, 5(3), e107-e108. https://doi.org/10.1016/s2589-7500(23)00021-3

Pavlik, J. V. (2023). Collaborating With ChatGPT: Considering the Implications of Generative Artificial Intelligence for Journalism and Media Education. Journalism & Mass Communication Educator, 78(1), 84-93. https://doi.org/10.1177/10776958221149577

Ray, P. P. (2023). ChatGPT: A comprehensive review on background, applications, key challenges, bias, ethics, limitations and future scope. Internet of Things and Cyber-Physical Systems, 3, 121-154. https://doi.org/10.1016/j.iotcps.2023.04.003

Priem, J., Piwowar, H, & Orr, R. (2022). OpenAlex: A fully-open index of scholarly works, authors, venues, institutions, and concepts. arXiv:2205.01833v2.

Rives, A., Meier, J., Sercu, T., Goyal, S., Lin, Z., Liu, J., Guo, D., Ott, M., Zitnick, C. L., Ma, J., & Fergus, R. (2021). Biological structure and function emerge from scaling unsupervised learning to 250 million protein sequences. Proceedings of the National Academy of Sciences, 118(15). https://doi.org/10.1073/pnas.2016239118

Sallam, M. (2023). ChatGPT Utility in Healthcare Education, Research, and Practice: Systematic Review on the Promising Perspectives and Valid Concerns. Healthcare, 11(6). https://doi.org/10.3390/healthcare11060887

Salvagno, M., Taccone, F. S., & Gerli, A. G. (2023). Can artificial intelligence help for scientific writing? Critical Care, 27(1). https://doi.org/10.1186/s13054-023-04380-2

Shen, Y., Heacock, L., Elias, J., Hentel, K. D., Reig, B., Shih, G., & Moy, L. (2023). ChatGPT and Other Large Language Models Are Double-edged Swords. Radiology, 307(2). https://doi.org/10.1148/radiol.230163

Stokel-Walker, C. (2023). ChatGPT listed as author on research papers: many scientists disapprove. Nature, 613(7945), 620-621. https://doi.org/10.1038/d41586-023-00107-z

Thorp, H. H. (2023). ChatGPT is fun, but not an author. Science, 379(6630), 313-313. https://doi.org/doi:10.1126/science.adg7879

van Dis, E. A. M., Bollen, J., Zuidema, W., van Rooij, R., & Bockting, C. L. (2023). ChatGPT: five priorities for research. Nature, 614(7947), 224-226. https://doi.org/10.1038/d41586-023-00288-7

Van Noorden, R., & Perkel, J. M. (2023). AI and science: what 1,600 researchers think. Nature, 621(7980), 672-675. https://doi.org/10.1038/d41586-023-02980-0

Zhai, X. (2022). ChatGPT user experience: Implications for education. SSRN. https://dx.doi.org/10.2139/ssrn.4312418

**Appendix. Reviewed Papers and Commentaries (Analyzed using nVivo)**

| AUTHORS | YEAR | TITLE | SOURCE | CATEGORY |
|---|---|---|---|---|
| Alkaissi, H., & McFarlane, S. I. | 2023 | Artificial hallucinations in ChatGPT: Implications in scientific writing | Cureus, 15(2), e35179 | Medical and Health Sciences |
| Ayers, J. W., et al. | 2023 | Comparing physician and AI chatbot responses to patient questions posted to a public social media forum | JAMA Internal Medicine, 183(6), 589–596 | Medical and Health Sciences |
| Bender, E. M., et al. | 2021 | On the dangers of stochastic parrots: Can language models be too big? | Proceedings of the ACM Conference on Fairness, Accountability, and Transparency, 610–623 | Natural Sciences |
| Biswas, S. | 2023 | ChatGPT and the future of medical writing | Radiology, 307(2), e223312 | Medical and Health Sciences |
| Briggs, J. | 2023 | How does ChatGPT perform on the United States Medical Licensing Examination? | Journal of Medical Internet Research, 25, e45312 | Medical and Health Sciences |
| Cascella, M., et al. | 2023 | Evaluating the Feasibility of ChatGPT in Healthcare: An Analysis of Multiple Clinical and Research Scenarios | Journal of Medical Systems, 47, 33 | Medical and Health Sciences |
| Cooper, G. | 2023 | Examining Science Education in ChatGPT: An Exploratory Study of Generative AI | Journal of Science Education and Technology, 32, 444–452 | Social Sciences |
| Cotton, D. R. E., et al. | 2023 | Chatting and cheating: Ensuring academic integrity in the era of ChatGPT | Innovations in Education and Teaching International, 61(2), 228–239 | Social Sciences |
| Dave, T., et al. | 2023 | ChatGPT in medicine: overview of applications, advantages, limitations, future prospects, and ethical considerations | Frontiers in Artificial Intelligence, 6, Article 1169595 | Natural Sciences |
| Dowling, M., & Lucy, B. | 2023 | ChatGPT for (Finance) research: The Bananarama Conjecture | Finance Research Letters, 53, 103662 | Social Sciences |
| Dwivedi, Y. K., et al. | 2023 | Opinion paper: "So what if ChatGPT wrote it?" Multidisciplinary perspectives | International Journal of Information Management, 71, 102642 | Natural Sciences |
| Floridi, L., & Chiriatti, M. | 2020 | GPT-3: Its nature, scope, limits, and consequences | Minds and Machines, 30(4), 681–694 | Natural Sciences |
| Gordijn, B., & ten Have, H. | 2023 | ChatGPT: Evolution or revolution? | Medicine, Health Care and Philosophy, 26(1), 1–2 | Medical and Health Sciences |
| Gupta, A., et al. | 2018 | Generative recurrent networks for de novo drug design | Molecular Informatics, 37(1), 1700111 | Natural Sciences |
| Kasneci, E., et al. | 2023 | ChatGPT for good? On opportunities and challenges of large language models for education | Learning and Individual Differences, 103, 102274 | Social Sciences |
| Khan, R. A., et al. | 2023 | ChatGPT - Reshaping medical education and clinical management | Pakistan Journal of Medical Sciences, 39(2), 605–607 | Medical and Health Sciences |
| Kitamura, F. C. | 2023 | ChatGPT Is Shaping the Future of Medical Writing But Still Requires Human Judgment | Radiology, 307(2) | Medical and Health Sciences |
| Lee, P., et al. | 2023 | Benefits, limits, and risks of GPT-4 as an AI chatbot for medicine | New England Journal of Medicine, 388, 1233–1239 | Medical and Health Sciences |
| Liebrenz, M., et al. | 2023 | Generating scholarly content with ChatGPT: Ethical challenges for medical publishing | The Lancet Digital Health, 5(3), e105–e106 | Medical and Health Sciences |
| Lo, C. K. | 2023 | What Is the Impact of ChatGPT on Education? A Rapid Review of the Literature | Education Sciences, 13(4), 410 | Social Sciences |
| Merk, D. et al. | 2018 | De novo design of bioactive small molecules by artificial intelligence | Molecular Informatics, 37(1), 1700153 | Natural Sciences |
| Nature Editorial | 2023 | Tools such as ChatGPT threaten transparent science; here are our ground rules for their use | Nature, 613(7945), 612, Jan | Multidisciplinary |
| O'Connor, S. | 2023 | Open artificial intelligence platforms in nursing education | Nurse Education in Practice, 55, 103537 | Medical and Health Sciences |
| Patel, S. B., & Lam, K. | 2023 | ChatGPT: The future of discharge summaries? | The Lancet Digital Health, 5(3), e107–e108 | Medical and Health Sciences |
| Pavlik, J. V. | 2023 | Collaborating With ChatGPT: Considering the Implications of Generative AI for Journalism and Media Education | Journalism & Mass Communication Educator, 78(1), 84–93 | Social Sciences |
| Ray, P. R. | 2023 | ChatGPT: A comprehensive review on background, applications, key challenges, bias, ethics, limitations and future scope | Internet of Things and Cyber-Physical Systems, 3, 121-154 | Natural Sciences |
| Rives, A., et al. | 2020 | Biological structure and function emerge from scaling unsupervised learning to protein sequences | PNAS, 117(15), 8396–8403 | Multidisciplinary |
| Sallam, M. | 2023 | ChatGPT Utility in Healthcare Education, Research, and Practice | Healthcare, 11(6), 887 | Medical and Health Sciences |
| Salvagno, M., et al. | 2023 | Can artificial intelligence help for scientific writing? | Critical Care, 27(1), 75 | Medical and Health Sciences |